\documentclass[reqno,10pt]{amsart}
\usepackage{hyperref}
\usepackage{graphicx}
\usepackage{slashed}
\usepackage{amscd}
\usepackage{tensor}
\usepackage{amssymb}
\usepackage{esint}
\usepackage{enumitem}
\usepackage{accents}
\usepackage[dvipsnames]{xcolor}
\usepackage[utf8]{inputenc}
\usepackage[off]{auto-pst-pdf}
\usepackage{pst-grad}
\usepackage{pst-plot}
\usepackage[mathscr]{eucal}
\usepackage{tagging}
\droptag{NotReady}

\numberwithin{equation}{section}
\allowdisplaybreaks[1]

\newtheorem{Def}{Definition}

\newtheorem{Example}[Def]{Example}

\newcommand{\beq}{\begin{equation}}
\newcommand{\eeq}{\end{equation}}
\newcommand{\Proof}{\begin{proof}}
\newcommand{\QED}{\end{proof} \noindent}
\newcommand{\QEDrem}{\ \hfill $\Diamond$}

\newcommand{\R}{\mathbb{R}}

\newcommand{\N}{\mathbb{N}}

\newcommand{\A}{\mathcal A}
\newcommand{\C}{\mathcal{C}}

\newcommand{\T}{{\mathcal{T}}}

\DeclareFontFamily{OT1}{rsfso}{}
\DeclareFontShape{OT1}{rsfso}{m}{n}{ <-7> rsfso5 <7-10> rsfso7 <10-> rsfso10}{}
\DeclareMathAlphabet{\mycal}{OT1}{rsfso}{m}{n}

\newcommand{\itemD}{\item[{\raisebox{0.125em}{\tiny $\blacktriangleright$}}]}
\newcommand{\iD}{\raisebox{0.125em}{\tiny $\blacktriangleright$}}

\newcommand{\bei}{\begin{itemize}[label=$\circ$,itemsep=.5em,leftmargin=*]}
\newcommand{\beii}{\begin{itemize}[label=$\rightarrow$,itemsep=.5em,topsep=.5em,leftmargin=*]}
\newcommand{\eni}{\end{itemize}}
\newcommand{\enii}{\end{itemize}}

\definecolor{darkblue}{RGB}{0,91,163}

\newcommand{\structure}{\mathbb S}
\newcommand{\domain}{\mathcal A}

\usepackage{phaistos}

\usepackage{mdframed} 

\newcommand{\ql}[1]{[\rotatebox{90}{\scalebox{0.25}{later}}]}

\newcommand{\cbb}[1]{   
\begin{itemize}[label=\iD,topsep=.5em,itemsep=.5em,leftmargin=*]
\item \textbf{#1}
    \begin{itemize}[label=-,itemsep=.2em,topsep=.2em,leftmargin=*]
}
\newcommand{\cbe}{      
    \end{itemize}
\end{itemize}
}

\usepackage[onehalfspacing]{setspace}

\usepackage{mdframed}
\definecolor{lightblue}{RGB}{225,244,249}			
\definecolor{lightgreen}{RGB}{218,247,232}          
\mdtheorem[style=Box,backgroundcolor=Bsp]{example}{Example}[Def]
\theoremstyle{plain}
\newmdenv[linewidth=0,needspace=130pt,skipabove=5pt,skipbelow=5pt,innertopmargin=10pt,
splittopskip=30pt,splitbottomskip=18pt,innerbottommargin=10pt,frametitleaboveskip=0pt,backgroundcolor=lightblue]{quick-example}
\newmdenv[linewidth=0,skipabove=5pt,skipbelow=5pt,innertopmargin=10pt,
splittopskip=30pt,splitbottomskip=18pt,innerbottommargin=10pt,frametitleaboveskip=0pt,backgroundcolor=lightblue]{defi}
\newmdenv[linewidth=0,skipabove=5pt,skipbelow=5pt,innertopmargin=10pt,
splittopskip=30pt,splitbottomskip=18pt,innerbottommargin=10pt,frametitleaboveskip=0pt,backgroundcolor=lightblue]{terminology}
\newmdenv[linewidth=0,skipabove=5pt,skipbelow=5pt,innertopmargin=10pt,
splittopskip=30pt,splitbottomskip=18pt,innerbottommargin=10pt,frametitleaboveskip=0pt,backgroundcolor=lightgreen]{reality}

\usepackage{cases}

\newcommand{\MDC}{\hyperlink{MDC}{(MDC)}}    
\newcommand{\MSC}{\hyperlink{MSC}{(MSC)}}    

\makeatletter
\def\section{\@startsection{section}{1}%
  \z@{.7\linespacing\@plus\linespacing}{.5\linespacing}%
  {\Large\scshape\centering}}
\makeatother

\begin{document}

\begin{center}
\Huge What is a Mathematical Structure\\of Conscious Experience?
\end{center}

\vspace{1cm}

\centerline{\Large Johannes Kleiner$^{1,2,3}$ and Tim Ludwig$^{4}$}
\vspace*{.3cm}
\centerline{$^1$Munich Center for Mathematical Philosophy, LMU Munich}
\centerline{$^2$Munich Graduate School of Systemic Neurosciences, LMU Munich}
\centerline{$^3$Association for Mathematical Consciousness Science}
\centerline{$^4$Institute for Theoretical Physics, Utrecht University}\vspace{-.3em}
\centerline{\small Princetonplein 5, 3584 CC Utrecht, The Netherlands}

\vspace{.7cm}

\begin{quote}
\textsc{Abstract.}
In consciousness science, several promising approaches have been developed for how to represent conscious experience in terms of mathematical spaces and structures.
What is missing, however, is an explicit definition of what a `mathematical structure of conscious experience' is. Here, we propose such a definition.
This definition provides a link between the abstract formal entities of mathematics and the concreta of conscious experience; it complements recent approaches that study quality spaces, qualia spaces or phenomenal spaces; it provides a general method to identify and investigate structures of conscious experience; and it may serve as a framework to unify the various approaches from different fields. We hope that ultimately this work provides a basis for developing a common formal language to study consciousness.

\end{quote}


\vspace{.7cm}

Attempts to represent conscious experiences mathematically
go back at least to 1860~\cite{fechner1966elements}, and a large number of approaches have been developed since. They span psycho-physics~\cite{klincewicz2011quality,kostic2012vagueness,kuehni2008color,renero2014consciousness,young2014quality}, philosophy~\cite{clark1996sensory,clark2000theory,coninx2022multidimensional,fortier2020multi,gert2017quality,lee2021modeling,lee2022objective,rosenthal2010think,rosenthal2015quality,rosenthal2016quality}, phenomenology~\cite{prentner2019consciousness,yoshimi2007mathematizing}, neuroscience~\cite{tallon2022topological,zaidi2013perceptual}, theories of consciousness~\cite{haun2019does,hoffman2014objects,mason2013consciousness,oizumi2014phenomenology} and mathematical consciousness science~\cite{grindrod2018human,kleiner2020mathematical,resende2022qualia,stanley1999qualia,tsuchiya2022enriched,tsuchiya2021relational}, and are
known under various different names, including quality spaces~\cite{clark1996sensory}, qualia spaces~\cite{stanley1999qualia}, experience spaces~\cite{kleiner2021falsification,kleiner2021mathematical}, Q-spaces~\cite{chalmers2022wave}, Q-structure~\cite{lyre2022neurophenomenal}, $\Phi$-structures~\cite{tononi2015integrated}, perceptual spaces~\cite{zaidi2013perceptual}, phenomenal spaces~\cite{fink2021structural}, spaces of subjective experience~\cite{tallon2022topological}, and spaces of states of conscious experiences~\cite{kleiner2020brain}.%
    \footnote{
    There is a vast literature of articles that either make use of these constructions or which offer important insights that concern these structures;
    for examples, see~\cite{ay2015information,barbosa2020measure,blum2021theory,blum2021theoretical,durham2020formal,ebner2022communication,grindrod2021cortex,hoffman2023fusions,jost2021information,kent2021beyond,kent2018quanta,kremnizer2015integrated,langer2022learning,lee2019microstructure,lee2022objective,mason2021model,rosas2020reconciling,rudrauf2017mathematical,seth2021predictive,signorelli2021reasoning,tsuchiya2021relational,tsuchiya2016using,tull2021integrated}.
    }
The mathematical structures and spaces, which these approaches introduced, have enabled important results in their respective fields. Yet, they remain largely fragmented. 
What is missing, from our perspective, is a definition of what the term `mathematical structure of conscious experience' refers to in the first place.

In this article, we propose a definition of mathematical structures of conscious experience.
Our main desideratum is that for a mathematical structure to be \textit{of} conscious experience, there must be something \emph{in} conscious experience that corresponds to that structure. We call this ``something'' a structural aspect of conscious experience.

Our key idea is to use variations to identify and investigate structural aspects of conscious experience. That is because the concept of variation can serve as a binding link between conscious experiences and mathematical structures: on the one hand, variations relate to conscious experiences, because they change some of their aspects (like qualia, qualities, or phenomenal properties); on the other hand, variations relate to mathematical structures, because they may or may not preserve them.

In defining what it means for a mathematical structure to be \textit{of} conscious experience, our proposal does not answer the question of what this mathematical structure is. Rather, it provides an analysandum for future work on spaces and structures of conscious experience.
Furthermore, by providing a general way to identify and investigate such structures, our proposal provides a framework to unify the various approaches from different fields.

Before presenting the details of our proposal, we discuss how recent approaches relate mathematical structures to conscious experience in Section~\ref{sec:previous}; we identify three key problems. 
In Section~\ref{sec:proposal}, we present our proposal together with the necessary background information. In Sections~\ref{sec:ex:relativeSimilarity}, \ref{sec:ex:metric}, and~\ref{sec:ex:topological}, we consider three important examples; namely relative similarity, metric spaces, and topological spaces. In Section~\ref{sec:threeProbsAgain}, we show how our proposal resolves the three problems identified in Section~\ref{sec:previous}.

\section{The Status Quo}\label{sec:previous}

So where do things stand? Most of the early work that has attributed mathematical structure to conscious experience was grounded in intuition. That is, whether or not a specific mathematical structure pertains to consciousness was not assessed systematically; instead, it was assessed based on an intuitive insight of appropriateness. 
More recent approaches have realized the need for a more systematic method, for example~\cite{gert2017quality,lee2021modeling,lee2023degrees,prentner2019consciousness,resende2022qualia,rosenthal2015quality,rosenthal2016quality}.
In this section, we analyze what we take to be the condition that underlies these approaches: a condition on what justifies prescribing a mathematical structure to conscious experience. As we will see, this condition is quite natural. But, as we will show, when understood as a sufficient condition, it is problematic.

In a nutshell, a mathematical structure consists of two building blocks; for a detailed introduction, see Section~\ref{sec:mathematicalStructure}. The first building block brings in one or more sets called the \emph{domains} of the structure. The second building block are \emph{relations or functions} which are defined on the domains. For reasons explained below, we will denote them as \emph{structures} in the narrow sense of the term. A metric space, for example, is a mathematical structure that is defined on the two domains: a set of points and the real numbers. Furthermore, it comprises a function---the so-called metric function---which maps two points to a real number.
A topological space, to give another example, is a mathematical structure that is defined on a single domain: a set of points. Furthermore, it comprises a collection of unary relations, which are subsets of the domain.%
    \footnote{A unary relation on a domain, in the mathematical sense, is a subset of the domain; see Section~\ref{sec:ex:topological}.}

Usually, a mathematical structure also comes with \emph{axioms}. The axioms establish conditions that the functions or relations have to satisfy. In the case of a metric structure, the axioms require the metric function to satisfy three conditions, called positive definiteness, symmetry, and triangle inequality. In the case of a topological structure, the axioms ensure the collection includes the empty set and the whole domain, that it is closed under finite intersections, and that it is closed under arbitrary unions.

When put in these terms, recent proposals that go beyond intuitive assessments, make use, either directly or indirectly, of the following condition to justify that a specific mathematical structure pertains to consciousness.
\hypertarget{MDC}{}
\begin{terminology}
\textbf{A mathematical structure describes conscious experience (MDC)} if and only if the following two conditions are satisfied:
\begin{enumerate}[label=(D\arabic*)]
    \item\label{MDC:domain} The domains of the structure 
    are sets whose elements correspond to aspects of conscious experiences.
    \item\label{MDC:axioms} The axioms of the structure are satisfied.
\end{enumerate}
\end{terminology}

\noindent Here, we use the term \emph{aspect} as a placeholder for qualia, qualities, (instantiated) phenomenal properties or similar concepts.

In the case of the metric structure introduced in~\cite{clark1996sensory}, for example, \ref{MDC:domain} is satisfied because the set of points corresponds to qualities of conscious experience. The real numbers might have a phenomenal interpretation as describing degrees of similarity, as for example in~\cite{lee2021modeling}; for details see Section~\ref{sec:ex:metric}. Condition~\ref{MDC:axioms} requires positive definiteness, symmetry, and the triangle inequality to hold. This includes, for example, the condition that ``points should have distance zero just in case the qualities represented by those points are phenomenally identical''~\cite[p.\,14]{lee2021modeling}.
In the case of the topological structure introduced in~\cite{stanley1999qualia}, to give another example, \ref{MDC:domain} is satisfied because the domain of the structure refers to qualia. Condition~\ref{MDC:axioms} would require, then, that the chosen collection of subsets satisfies the axioms of a topological space.

Prima facie,~\MDC\ could be taken to define what a mathematical structure of conscious experience is. However, if understood as sufficient condition, three problems arise.
This implies, in particular, that~\MDC\ cannot be used to justify that a mathematical structure pertains to consciousness.

\subsection*{Problem 1: Incompatible Structures} 
A first reason why~\MDC\ cannot be a sufficient condition to asses whether a mathematical structure pertains to consciousness is that it would allow for incompatible structures to pertain to consciousness.

Consider, as an example, the case of topology. A basic question in topology is whether a target domain is discrete or not. A target domain is discrete if and only if its topology contains all subsets of the domain~\cite{joshi1983introduction}. Otherwise, the target domain is not discrete. 
These two cases are exclusive, meaning that discrete and non-discrete topological structures are incompatible.

According to~\MDC, conscious experience has a discrete structure. That is because any set whatsoever can be equipped with the discrete topology. Therefore, picking a set $X$ of aspects (qualia, qualities, phenomenal properties, etc.) and choosing its discrete topology provides a mathematical structure that satisfies both conditions~\ref{MDC:domain} and~\ref{MDC:axioms}. But, according to~\MDC, consciousness also has a non-discrete structure. That is because any set can also be equipped with a non-discrete topology. We can, for example, take an arbitrary decomposition of the set $X$ into two subsets $A$ and $A^\perp$, where $A^\perp$ is the complement of $A$, and consider the topology $\{\emptyset,A,A^\perp,X\}$. This choice satisfies all axioms of a topology, and therefore satisfies~\ref{MDC:axioms}. Furthermore, it is built on the same set $X$ as the discrete topology above, which implies that it also satisfies~\ref{MDC:domain}. Therefore, the discrete and the non-discrete topological structures are both structures of conscious experience, according to~\MDC.

This example shows that, if understood as sufficient condition,~\MDC\ implies that two incompatible structures pertain to conscious experience and that they do so with respect to the exact same domain of aspects. The condition fails to determine which of the two incompatible structures is the right one, or to remain silent on the issue.

\subsection*{Problem 2: Arbitrary Re-Definitions.} A second reason why~\MDC\ cannot be a sufficient condition is that it allows for arbitrary re-definitions:
if one structure is given that satisfies~\MDC, then any arbitrary definition of a new structure in terms of the given structure also satisfies~\MDC, so long as the domains of the structure remain unchanged. If the former pertains to consciousness, so does the latter.

A simple and well-behaved example of this is given by rescaling a metric function. Let us suppose that $(M,d)$ is a metric structure which pertains to consciousness according to~\MDC, where $M$ is a set of aspects and $d$ is the metric function, which provides a real number $d(a,b)$ for every two aspects $a$ and $b$. Since $(M,d)$ satisfies~\MDC, so does every structure $(M, C\cdot d)$, where $C\cdot d$ is the multiplication of the function $d$ by a positive real number~$C$. Here, the number~$C$ can be chosen arbitrarily. If one metric structure pertains to consciousness according~\MDC, so does an uncountably infinite number of metric structures. 

What is more, when re-defining structures, one is free to change the axioms as one pleases. For example, we could pick any function $f$ that maps $M$ to the positive real numbers and define a new distance function by $(f(a)+f(b)) \cdot d(a,b)$. This is not a metric structure anymore, because the triangle inequality axiom does not hold.
But it still satisfies positive definiteness and symmetry, and therefore satisfies~\MDC, with a new set of axioms. One could even break asymmetry to get a distance function like the one applied by IIT~\cite{kleiner2021mathematical}. More severe cases appear with more complicated structures.

This is a problem, not only because of the unlimited number of structures that appear, but also because there is an arbitrariness in the definition of a new structure, specifically concerning the axioms. It seems strange that the axioms can be redefined at will, so as to always satisfy Condition~\ref{MDC:axioms}. Something is missing that restricts this arbitrariness in~\MDC.

\subsection*{Problem 3: Indifference to Consciousness.} The third reason, which speaks against the sufficiency of~\MDC, is that the proposed condition seems somewhat indifferent to details of conscious experience. 

To illustrate this indifference, let us consider again the discrete and non-discrete topological structures from above. As we have shown, these structures pertain to conscious experience according to~\MDC. Yet, nothing more than a few lines needed to be said to establish this fact. In particular, we did not need to use any noteworthy input related to consciousness other than picking some set of aspects; and it didn't matter which aspects we picked.

It is a red flag if so short an analysis, which does not depend on consciousness in a meaningful way, establishes facts about the mathematical structure of conscious experience. This is another hint that condition~\MDC\ is  missing some important piece, if used as sufficient condition.

\subsection*{The Way Forward}
To resolve the three problems, our task is to find the missing piece and to propose a definition for a mathematical structure of conscious experience that makes sense as a necessary and sufficient condition. Two desiderata guide our search. First, 
as is the case with~\MDC, the definition should be \emph{about} conscious experience in the sense that it describes aspects of conscious experience. Second, there should be something \emph{in} conscious experience that relates to a mathematical structure if that structure is a mathematical structure \emph{of} conscious experience. This ``something'' should make sure that the definition is not indifferent to conscious experience (Problem~3) and that it relates to the mathematical structure in a meaningful way, so as to stop arbitrary re-definitions (Problem~2). The proposal which we present in the next section is the result of our search.

Looking back at Condition~\MDC\ after our analysis, we think that~\MDC\ is best 
understood as an expression of what it takes for a mathematical structure to \emph{describe} conscious experience. Because of the problems with sufficiency, a structure that satisfies this condition might not pertain to consciousness; but it might still be a valuable descriptive tool that is distinguished from other structures by its relation to aspects of conscious experience. This is why, retrospectively, we have chosen the term `describes conscious experiences' when specifying the condition. Our first desideratum will lead us to develop a new condition that contains~\MDC\ as necessary part; this is aligned with the intuition that any mathematical structure of conscious experience also describes conscious experience.

\section{Mathematical Structures of Conscious Experience}\label{sec:proposal}

For a mathematical structure to be \emph{of} conscious experience, rather than just a descriptive tool \emph{for} conscious experience, there should be something \emph{in} conscious experience that relates to that structure. Denoting a mathematical structure by $S$, we call this  structural aspect of consciousness an $S$-aspect.

To make sense of what an $S$-aspect is, we need to understand how aspects (like qualia, qualities or phenomenal properties) relate to mathematical structures. While aspects may have an arity (meaning they may be instantiated \emph{relative to other aspects}), they are not experienced as having a mathematical structure per se.%
    \footnote{With the exception of experiences of mathematical structures themselves, of course.}
Therefore, relating aspects to mathematical structures 
requires a tool that applies both; concreta and abstract formal entities.
Variations provide such a tool.

In general, a variation is a change of something into something else; in our case, it is a change of one experience into another experience. Such variations may be induced by external stimuli or interventions, come about naturally, or be subjected to imagination (`imaginary variations'~\cite{husserl1936crisis}). 
Variations are intimately related to aspects of conscious experiences because they may or may not change them. And they are intimately related to mathematical structures, because they may or may not preserve them. An $S$-aspect, then, is an aspect that is changed by a variation if and only if the variation does not preserve the structure $S$. To explain this in detail is the purpose of the remainder of this section.

\subsection{Terminology and Notation}\label{sec:basics}

Here, we introduce the key terms we use to define mathematical structures of conscious experience.
These terms are \emph{conscious experiences}, \emph{aspects} of conscious experiences, and \emph{variations} of conscious experiences. The introduction proceeds axiomatically, so that our construction does not rely on a specific choice of these concepts. Rather, any choice that is compatible with what we say here can be the basis of an application of our definition.

Our construction is based on a \emph{set $E$ of conscious experiences} of an experiencing subject.
We denote individual conscious experiences in that set by symbols like $e$ and $e'$; formally $e, e' \in E$. From a theoretical or philosophical perspective, one may think of the set $E$ as comprising all conscious experiences which one experiencing subject can have, i.e. all nomologically possible experiences of that subject. From an experimental or phenomenological perspective, one may think of this set as comprising all conscious experiences that can be induced in the lab or in introspection. Different such choices may lead to different mathematical structures being accessible.

We use the term \emph{aspect} as a placeholder for concepts such as \emph{qualia}~\cite{tye2021qualia}, \emph{qualities}~\cite{rosenthal2010think}, or (instantiated) \emph{phenomenal properties}.%
    \footnote{Many other concepts work as well. For example, if one works with an atomistic conception of states of consciousness, where the total phenomenal state of a subject---what it is like to be that subject at a particular time---is built up from individual atomic states of consciousness, one can take $e$ to denote the total phenomenal state and aspects to be the \emph{states of consciousness} in that total state. Another example would be to take aspects to denote phenomenal distinctions as used in Integrated Information Theory~\cite{tononi2015integrated}.
    What matters for our definition to be applicable is only that according to one's chosen concept of conscious experience, every conscious experience exhibits a set of aspects.
    }
For every experience $e \in E$, we denote the set of aspects instantiated in this experience by $A(e)$. The set of {all aspects} of the experiences in $E$, denoted by $\A$, is the union of all $A(e)$; formally $\A = \bigcup_{e \in E} A(e)$. 
Individual aspects, that is members of $\A$, will be denoted by small letters such as $a, b, c$. 
When explaining examples, we will often use the abbreviation `$a$ is the experience of ...' as a shorthand for saying `$a$ is a ... aspect of an experience'. For example, `$a$ is the experience of red color' means `$a$ is a red color aspect of an experience'.

Some aspects may require other aspects for their instantiation. For example, it is usually the case that an experience of relative similarity is an experience of relative similarity of something, for example two color aspects relative to a third color aspect. If an aspect $a$ requires other aspects for its instantiation, we will say that the aspect $a$ \emph{is instantiated relative to} aspects $b_1, ..., b_m$, or simply that $a$ \emph{is relative to} $b_1, ..., b_m$. Aspects which are instantiated relative to other aspects are the building blocks for the structure of conscious experience.

A \emph{variation} of a conscious experience $e$ changes $e$ into another experience $e'$. 
Because experiences have structure, there may be various different ways to go from $e$ to $e'$.%
    \footnote{To illustrate this point, consider, for example, the following two mappings $v$ and $v'$ which map the numbers $1$, $2$, and $3$ to the numbers $2$, $4$, and $6$. The mapping $v$ is the multiplication of every number by $2$, meaning that we have $v(1)=2$, $v(2)=4$, $v(3)=6$. The mapping $v'$, on the other hand, is defined by $v(1)=6$, $v(2)=2$, $v(3)=4$. If we only cared about the \emph{sets} of elements that these mappings connect, the mappings would be equivalent: there is no difference between the set $\{2,4,6\}$, which is the image of $v$, and $\{6,2,4\}$, which is the image of $v'$. If, however, we care about the \emph{ordering} of the elements of the sets, which we usually do in the case of numbers, then there is a difference. While $2 \leq 4 \leq 6$, it is not the case that $6 \leq 2 \leq 4$. Because we care about the order of the elements, we need to say which element goes where.
    }
Therefore, in addition to specifying $e$ and $e'$, a variation is a partial mapping
$$
v: A(e) \rightarrow A(e') \:.
$$
This mapping describes how aspects are replaced or reshuffled by the variation. A mapping which is not surjective, meaning that it does not map to all aspects in $A(e')$, makes room for appearance of new aspects. A mapping which is partial, meaning that it does not specify a target for every aspect in $A(e)$, makes room for aspects to disappear.

\subsection{What is a Mathematical Structure?}\label{sec:mathematicalStructure}
To find a rigorous definition of the mathematical structure of conscious experience, we need to work with a rigorous definition of mathematical structure. But, what is a mathematical structure? Fortunately, mathematical logic provides us with an answer to just that question.

A \emph{mathematical structure} $\structure$ consists of two things: domains, on the one hand, and functions or relations, on the other hand. We now introduce these concepts based on two simple examples.

The \emph{domains} of a structure $\structure$ are the sets on which the structure is built. We denote them by $\domain_i$, where $i$ is some index in a parameter range $I$. In the case of a metric structure, for example, the domains would be $\domain_1 = M$ and $\domain_2 = \R$, where $M$ is a set of points and $\R$ denotes the real numbers, understood as a set. In the case of a strict partial order, there is just one domain $\domain$, which contains the elements that are to be ordered.

The second ingredient are functions and/or relations. \emph{Functions} $f$ map some of the domains to other domains. In the case of a metric structure, the function would be a metric function
$d: M \times M \rightarrow \R$, which maps from $\domain_1 \times \domain_1$ to $\domain_2$.
A \emph{relation} $R$, in the mathematical sense, is a subset of the $m$-fold product $\domain_i \times ... \times \domain_i$. Here, $\domain_i$ is the domain on which the relation is defined, and $m$ is the arity of the relation, which expresses how many relata the relation relates. The product is usually just written as $\domain_i^m$. In the case of a strict partial order, the relation is binary, which means that $R$ is a subset of $\domain^2$. For binary relations, one usually uses notation like $a < b$ instead of writing $(a,b) \in R$; still, it is important to keep in mind that relations are subsets in that sense.

In almost all cases, mathematical structures also come with \emph{axioms}, which establish conditions that the functions or relations have to satisfy. They are useful because they constrain and classify the structure at hand. For $\structure$ to be a metric structure, for example, the function $d$ has to satisfy the axioms of positive definiteness, symmetry, and triangle inequality~\cite{rudin1976principles}. For $\structure$ to be a strict partial order, the relation $R$ has to be irrefelxive, asymmetric, and transitive~\cite{joshi1989foundations}. 

To have a nice and compact notation, we will use one symbol~$S_j$ to denote both functions relations. That is because, in any concrete proposal, it is always clear whether $S_j$ is a function or a relation.%
    \footnote{In mathematical logic, mathematical structures are denoted as triples of domains, relations, and functions.
    However, in our case, using just one symbol for functions and relations improves readability substantially.}
The index $j$ takes values in some parameter range $J$ that specifies how many functions or relations there are. In summary, the desired rigorous definition is:

\begin{terminology}
\textbf{A mathematical structure $\structure$} is a tuple
$$ 
    \structure = \big( (\domain_i)_{i \in I}, (S_j)_{j \in J} \big)
$$ 
of domains $\domain_i$ and functions or relations $S_j$.
\end{terminology}

For given domains $\A_i$, the mathematical structure $\structure$ is fully determined by the $S_j$.
Thus, we can also refer to $S_j$ as `structures', if the domains are clear from context.
For simplicity, we can drop the index $j$ and simply write $S$ whenever we consider just one such structure.

As a final step in this section, we introduce the \emph{relata} of a structure $S$. This will be helpful to write things concisely below.
The term relata designates those elements that are related by a structure. In the case where~$S$ is a relation $R$ on a domain $\domain$ and has arity $m$, these are the elements of the $m$-tuples $(b_1, ..., b_m) \in R$.  In the case where $S$ is a function $f: \domain_1 \times ... \times \domain_{m-1} \rightarrow \domain_m$, the relata are the elements of the $m$-tuples $(b_1, ..., b_{m-1}, b_m)$ where $b_m = f(b_1, ..., b_{m-1})$, and where the other $b_i$ range over their whole domains. For notational simplicity, we write $b_1, ..., b_m$ instead of $(b_1, ..., b_m)$ when designating relata below.

\subsection{What is a Mathematical Structure of Conscious Experience?}\label{subsec:msc}
Finally, to the heart of the matter!
We recall that we have so far identified two desiderata for a mathematical structure $\structure$ to be a mathematical structure of conscious experience. First, it should be \emph{about} conscious experiences in the sense that it describes aspects of conscious experiences. Second, there should be aspects \emph{in} conscious experience that relate to the structure~$\structure$. The following definition satisfies these two desiderata.

\hypertarget{MSC}{}
\begin{reality}\label{def:structure_of_consc}
\textbf{A mathematical structure $\structure$ is a mathematical structure of conscious experience (MSC)} if and only if the following two conditions hold:
\begin{enumerate}[label=(S\arabic*)]
\item The domains $\domain_i$ of $\structure$ are subsets of $\A$.\label{MSC:domains}
\item For every $S_j$, there is a $S_j$-aspect in $\A$.\label{MSC:aspect}
\end{enumerate}
Here, $\A$ denotes the set of all aspects of the experiences in $E$; formally $\A = \bigcup_{e \in E} A(e)$, the $\domain_i$ denote the domains of the structure $\structure$, and the $S_j$-aspects are defined below.
\end{reality}

\noindent Condition~\ref{MSC:domains} guarantees that the first desideratum is satisfied. Condition~\ref{MSC:aspect} guarantees that the second desideratum is satisfied. Furthermore, whenever a certain \emph{type} of structure (metric, topological, partial order, manifold, etc.) is claimed to be a structure of conscious experience, the axioms that constrain and classify that type have to hold. Therefore, any mathematical structure of conscious experience~\MSC\ is also a mathematical structure that describes conscious experience according to~\MDC. The requirement that has been applied in previous proposals remains a necessary condition.

The remaining task of this section, then, is to explain what an $S_j$-aspect is.
For notational simplicity, we use the symbol $S$ to denote $S_j$.
As we have emphasized before, variations are key to understand the structure of conscious experience, because they link aspects and structure. Therefore, to be able to precisely define what an $S$-aspect is, we need to understand how variations relate to aspects, on the one hand, and structures, on the other hand.
Our strategy is to first discuss how variations relate to aspects. This amounts to specifying what precisely it means for a variation to change an aspect. Second, we focus on how variations relate to mathematical structure. This amounts to explaining what it means for a variation to preserve a structure. Finally, combing these two steps allows us to understand $S$-aspects
and provide a useful definition.

What does it mean for a variation $v:A(e) \rightarrow A(e')$ to change aspects?
The underlying idea is simply that an aspect is present in the source of the variation, $A(e)$, but not present any more in the target of the variation, $A(e')$.
We need to take into account, though, that aspects are often instantiated relative to other aspects (see Section~\ref{sec:basics}). This can be done as follows.

\begin{terminology}
    \textbf{A variation} $v: A(e) \rightarrow A(e')$ \textbf{changes an aspect} $a \in A(e)$ \emph{relative to} $b_1, ..., b_m \in A(e)$ if and only if $a$ is instantiated relative to $b_1, ..., b_m$ in $A(e)$, but $a$ is not instantiated relative to $v(b_1), ..., v(b_m)$ in $A(e')$.%
\end{terminology}

\noindent In the case where $a \in A(e)$ is not instantiated relative to other aspects, the definition indeed reduces to the simple condition that $a \in A(e)$ but $a \not \in A(e')$.
The negation of the definition is also as intuitively expected: the aspect is present both in the source and in the target.%
    \footnote{Because the definiendum already includes the first part of the condition, the negation is as follows:\\
    A variation $v: A(e) \rightarrow A(e')$ \textit{does not change an aspect} $a \in A(e)$ \textit{relative to} $b_1, ..., b_m \in A(e)$ if and only if $a$ is instantiated relative to $b_1, ..., b_m$ in $A(e)$ and $a$ is also instantiated relative to $v(b_1), ..., v(b_m)$ in $A(e')$.\\
    We felt that is the best way of writing things to optimize clarity.
    }

For applications it is important to understand that this definition can fail to apply in two ways. 
First, it can fail because there is no $a$ in $A(e')$ which is instantiated relative to $v(b_1), ..., v(b_m)$. This, in turn, can be the case either because there is not $a$ in $A(e')$ at all, or because there is an $a$ in $A(e')$ but it is instantiated relative to other aspects. Second, it can fail because one or more of the $v(b_1), ..., v(b_m)$ do not exist. The second case is possible because $v$ is a \emph{partial} mapping, which means aspects can disappear.

What does it mean for a variation to preserve a mathematical structure?
The underlying idea is that a variation preserves the structure if and only if the structure is satisfied before the variation and remains to be satisfied after the variation.
By its very nature, this is a mathematical condition, namely the condition of being a homomorphism, as specified by mathematical logic~\cite{mileti2022modern}.
The definition of a homomorphism, though, always applies to all elements of a domain at once. For our case, it is best to refine this definition to a single set of relata.%
    \footnote{For notational simplicity, we write $R\big(b_1,...,b_m\big) = R\big(v(b_1),...,v(b_m)\big)$ instead of 
    $R\big(b_1,...,b_m\big) \Leftrightarrow R\big(v(b_1),...,v(b_m)\big)$.
    }

\begin{terminology}
    \textbf{A variation} $v: A(e) \rightarrow A(e')$ \textbf{preserves a structure} $S$ 
    \emph{with respect to relata} $b_1, ..., b_m \in A(e)$ if and only if
    we have 
    \begin{enumerate}[label=(P\arabic*)]
        \item $R\big(b_1,...,b_m\big) = R\big(v(b_1),...,v(b_m)\big)$  if $S$ is a relation $R$, or\label{preserves:relation}
        \item $v\big( f(b_1, ..., b_{m-1}) \big) = f\big( v(b_1), ..., v(b_{m-1}) \big)$ 
        if $S$ is a function $f$.\label{preserves:function}
    \end{enumerate}
\end{terminology}

\noindent As in the previous case, the negation of this definition is exactly what is intuitively expected: a variation does not preserve the structure if and only if the structure is satisfied before the variation, but not satisfied after the variation.%
    \footnote{\textit{A variation} $v: A(e) \rightarrow A(e')$ \textit{does not preserve a structure} $S$ \emph{with respect to relata} $b_1, ..., b_m \in A(e)$ if and only if
    we have $R\big(b_1,...,b_m\big) \neq R\big(v(b_1),...,v(b_m)\big)$ if $S$ is a relation $R$, or $v\big( f(b_1, ..., b_{m-1} \big) \neq f\big( v(b_1), ..., v(b_{m-1}) \big)$ if $S$ is a function $f$.\\
    This negation agrees with the intuition because the definiendum already states part of the condition that follows, namely that $b_1, ..., b_m$ are relata of the structure $S$ in $A(e)$, which implies that $(b_1,...,b_m) \in R$ if $S$ is a relation and that $f(b_1, ..., b_{m-1})$ exists in $A(e)$ if $S$ is a function, meaning that the structure is satisfied before the variation.}

For applications it is again important to see that the definition can fail for two reasons. First, it could be the case that one or more of the $v(b_i)$ do not exist in $A(e')$, if the corresponding aspect disappears. Second, the identities may fail to hold.

We finally have the keys to understand $S$-aspects and provide a useful definition.
The underlying idea is that an $S$-aspect is an aspect that, under any variation, behaves exactly as the structure $S$ does. Whenever $S$ is preserved, the $S$-aspect does not change. 
Whenever the $S$-aspect changes, the structure $S$ is not preserved. That is, it needs to satisfy the following definition.
\begin{reality}
    \textbf{An aspect $a \in \A$ is a $S$-aspect} if and only if the following condition holds:\\
    A variation \emph{does not preserve} $S$ with respect to relata $b_1, ..., b_m$ \emph{if and only if} the variation \emph{changes} $a$ relative to $b_1, ..., b_m$.
\end{reality}

\noindent Here, the condition needs to hold true for all variations and all relata. This means that it needs to hold true for all variations of all experiences $e$ in the set $E$ that instantiate relata of the structure~$S$. 

This concludes our proposal for the definition of the mathematical structure of conscious experience.
It is a structure whose domains correspond to sets of aspects, and which contains an $S$-aspect for every relation or function of the structure. In the next three sections, we apply this definition to three examples. On the one hand, these examples illustrate the definition. On the other hand, they 
provide new insights to structures that have been featured prominently in previous approaches.

\section{Relative Similarity}\label{sec:ex:relativeSimilarity}

Our first example concerns relative similarity, which plays an important role, for example, in the construction of quality spaces by Austen Clark~\cite{clark1996sensory,clark2000theory}. In this example, we use natural language to pick out experiences and aspects. As we will see, this works fine to a large extend, but at one point we will have to show a bit of good faith when it comes to the precision of natural language.

\begin{figure}[t]
\includegraphics[width=10cm]{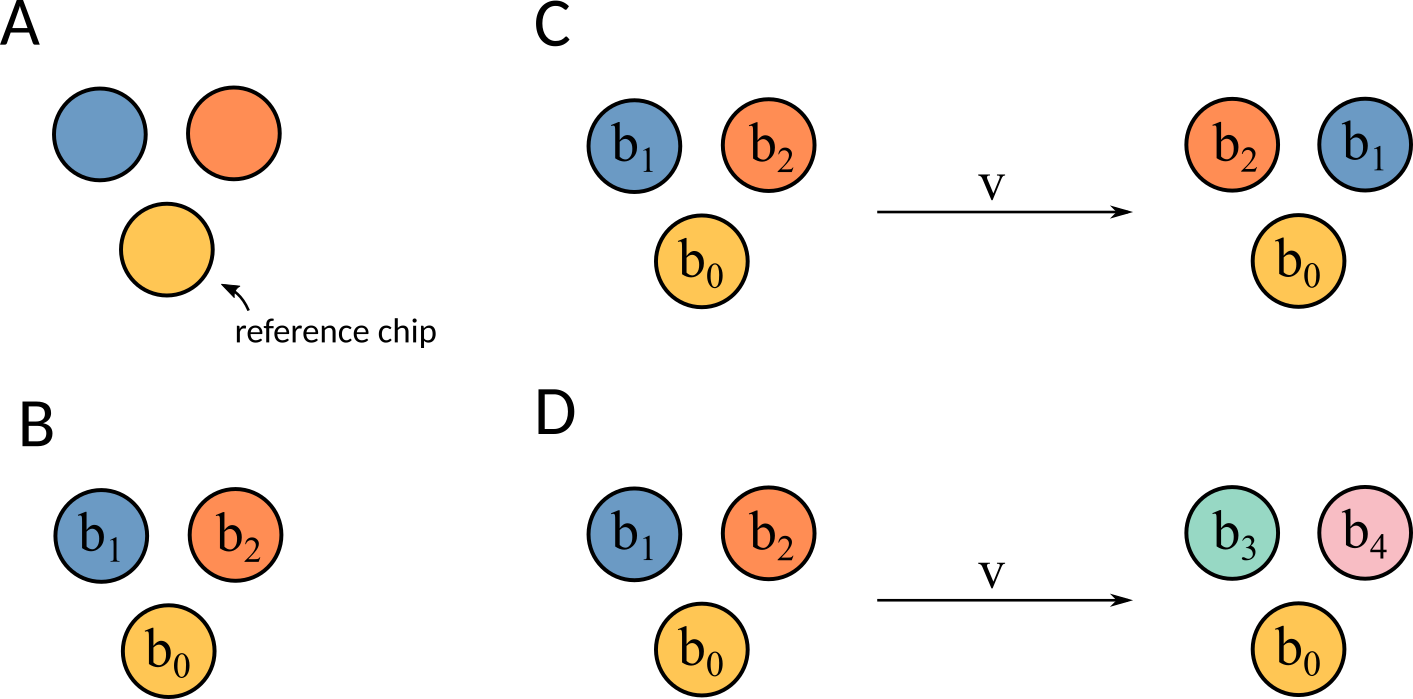}
\caption{To help explain the example of relative similarity, this figure illustrates experiences with color qualities and variations thereof. Subfigure~A illustrates an experience of three color chips as well as the concept of \emph{relative similarity}: many readers will experience the color of the top-left color chip to be \emph{less similar} to the reference chip than the color of the top-right color chip. Subfigure B illustrates our notation for the color aspects corresponding to the color chips. Subfigures C and D illustrate variations $v$ of experiences: a swap of two color aspects in C; and a replacement of two color aspects in D.}
\label{fig:colors}
\end{figure}

A first step in applying our definition is to choose a set $E$. Here we take $E$ to comprise experiences of three color chips, as indicated in 
Figure~\ref{fig:colors}A, where one of the chip (the reference) has a fixed color coating and the others vary in a range of color coatings $\Lambda$.\footnote{The ``color coating'' here denotes the physical stimuli. Alternative choices would be to speak of wavelength mixtures, presentation, etc.
} 

The second step is to specify the set of aspects $A(e)$ for every experience $e \in E$. Here, we take $A(e)$ to comprise:%
    \footnote{The experience $e$ may contain many other aspects. However, in $A(e)$ we only include those which are relevant for our investigation.}
(a) the color qualities in $e$, that is, the experienced colors of the individual chips; (b) positional qualities of the color experiences, that is, which chip has which color; and (c) the experience of \emph{relative similarity}. Relative similarity is an experience of one pair of aspects to be more, less, or equally similar to each other than another pair of aspects; here, the two pairs have to have one aspect---the reference---in common. In Figure~\ref{fig:colors}A, for example, the color of the top left chip will, for many readers, be less similar to the reference chip than the color of the top right chip.

To pick out relative similarity more precisely, we let $b_0$, $b_1$ and $b_2$ denote the color aspects of the three chips in an experience $e$, where $b_0$ is the color aspect of the reference; see Figure~\ref{fig:colors}B. For some experience $e$, it might be the case that the colors $b_1$ and $b_0$ are experienced as less similar to each other than the colors $b_2$ and $b_0$.
In this case, the experience $e$ has a relative similarity aspect in the above sense; we denote this ``less-similar'' relative similarity aspect by~$a$. So, $a$ is an aspect of $e$, and it is instantiated relative to $b_1$ and $b_2$.\footnote{To be precise, $a$ is also relative to $b_0$. But since $b_0$ does not vary in $E$ we can leave this implicit.}

Variations change one experience $e$ into another experience $e'$. An example for a variation would be a swap of the coatings of the two non-reference chips, as in Figure~\ref{fig:colors}C. Another example for a variation would be to change the coatings of both non-reference chips to some other coating in $\Lambda$, as in Figure~\ref{fig:colors}D. Formally, variations are represented by mappings $v: A(e) \rightarrow A(e')$. In the first example, Figure~\ref{fig:colors}C, the mapping is of the form $v(b_1) = b_2$ and $v(b_2) = b_1$, and $v(c) = c$ for all other aspects $c$, except for the relative similarity aspect $a$, which is discussed in detail below. In the second example, Figure~\ref{fig:colors}D, the mapping is as in the first example but with $v(b_1) = b_3$ and $v(b_2) = b_4$.

The key question of this example is: Is there a mathematical structure of conscious experience which corresponds to relative similarity? To answer this question, we propose a structure and check whether~\MSC\ applies.

The words ``less similar than'' in the description of relative similarity already indicate that some order, in the mathematical sense of the word, might be involved. For reasons that will become clear below, we propose a strict partial order as mathematical structure. A strict partial order $(\C,<)$, consists of a set $\C$, which is the domain of the structure, and a binary relation `$<$' on $\C$. For all $x,y,z \in \C$, this binary relation has to satisfy the following axioms:
\begin{itemize}
    \itemD \emph{Irreflexivity}, meaning that there is no $x \in \C$ with $x<x$.
    \itemD \emph{Asymmetry}, meaning that if $x<y$, then it is not the case that $y<x$.
    \itemD \emph{Transitivity}, meaning that if $x<y$ and $y<z$, then also $x<z$.
\end{itemize}

In order to propose a strict partial order structure of conscious experiences, we need to specify how the set $\C$ and the relation $<$ relate to (aspects of) conscious experience.
For the set $\C$ we choose the color qualities of the experiences in $E$, meaning that $\C$ now comprises the color qualities evoked by the coatings $\Lambda$ of the chips we consider. For example, it contains what we have labelled $b_0$, $b_1$, $b_2$, $b_3$ and $b_4$ in Figure~\ref{fig:colors}. For the relation, we define $b_i < b_j$ if and only if $b_i$ is experienced as less similar to $b_0$ than $b_j$ is to $b_0$.%
    \footnote{Since relative similarity, as defined above, depends on the choice of reference $b_0$, it would be more precise to write $<_{b_0}$ instead of $<$. However, to simplify the notation, we keep the reference implicit.}

For this proposal to make sense, we first need to check whether the axioms are satisfied. If they were not satisfied, the proposal could still be a structure of conscious experience; but it wouldn't be a strict partial order. That's why the axioms are not explicitly mentioned in~\MSC. Irreflexivity is satisfied because no color quality is less similar to the reference than itself. Asymmetry is satisfied because if a $b_i$ is less similar to the reference than $b_j$, then $b_j$ is not less similar to the reference than $b_i$.

Transitivity is a more interesting case. The use of terms like `less similar to' in natural language suggests that transitivity is also satisfied; it suggests that, if $b_i$ is less similar to the reference than $b_j$ and $b_j$ is less similar to the reference than $b_k$, then $b_i$ should be less similar to the reference than $b_k$. But it might very well be the case that natural language is not precise enough to describe its target domain. The use of natural language may be justified in simple cases, or even in a majority of cases, but whether or not transitivity holds for all $b_i,b_j,b_k \in \C$ is, ultimately, an empirical question. Already in such a simple example one can see how the mathematical-structure approach might be used to identify subtle details of conscious experiences for which natural language alone is not precise enough. Still, for the purpose of this example, we're going to show the above-mentioned bit of good faith and assume transitivity to hold as well.

Having checked that the axioms hold---that is, that the proposal is indeed a strict partial order---we can proceed to check whether the structure is a mathematical structure of conscious experience according to \MSC. Concerning Condition~\ref{MSC:domains}, there is one domain $\C$ and it consists of color qualities, so this condition is satisfied. Therefore, only Condition~\ref{MSC:aspect} remains to be checked.

We claim that the relative similarity aspect $a$, as defined above, is in fact a $<$-aspect. To see that this is true, we have to show that a variation does not preserve $<$ with respect to relata $b_1$ and $b_2$ if and only if the variation changes $a$ relative to $b_1$ and $b_2$.

Consider any variation $v: A(e) \rightarrow A(e')$ that does not preserve $<$ with respect to relata $b_1, b_2 \in A(e)$. Two aspects $b_1$ and $b_2$ are relata of $<$ if either $b_1 < b_2$ or $b_2 < b_1$. We focus on the first case as the other one follows from the first by renaming $b_2$ and $b_1$ in what follows.
By definition of the $<$ relation, $b_1 < b_2$ means that $b_1$ is experienced as less similar to the reference than $b_2$. Therefore, there is also a relative similarity aspect $a \in A(e)$ as defined above.
As explained in Section~\ref{subsec:msc}, there can be two ways in which the variation $v$ might not preserve $<$. Either $v(b_1)$ or $v(b_2)$ are not defined, or, if they are defined, it is not the case that $v(b_1) < v(b_2)$. In the former case, there cannot be an $a$ in $A(e')$ relative to $v(b_1)$ or $v(b_2)$, simply because the latter do not both exist. In the latter case, it follows from the definition of $<$ that $v(b_1)$ is not experienced as less similar to the reference than $v(b_2)$. So, there is no $a \in A(e')$ relative to $v(b_1)$ and $v(b_2)$. Hence, we may conclude that $v$ changes $a$ relative to $b_1$ and $b_2$.

For the opposite case, let $v: A(e) \rightarrow A(e')$ be a variation which preserves $<$ with respect to relata $b_1$ and $b_2$. As before, this implies that $a$ is in $A(e)$ relative to $b_1$ and $b_2$. Because $v$ preserves $<$, $v(b_1)$ and $v(b_2)$ both exist and we also have $v(b_1) < v(b_2)$. Applying the definition of $<$ then implies that $a$ is also in $A(e')$ relative to $v(b_1)$ and $v(b_2)$. Hence $v$ does not change~$a$ relative to $b_1$ and $b_2$.

Because in both of these cases, $v$ was arbitrary, it follows that $a$ is indeed a $<$-aspect. Therefore, Conditions~\ref{MSC:domains} and~\ref{MSC:aspect} of~\MSC\ are both satisfied, and the strict partial order $(\C,<)$ is indeed a mathematical structure of conscious experience; it is the mathematical structure of relative similarity of color experiences with respect to $b_0$.

\section{Metric Structure}\label{sec:ex:metric}

Our next example is a metric structure. Metric structures feature prominently in various different approaches, for example~\cite{clark1996sensory,clark2000theory,lee2021modeling,rosenthal2010think,rosenthal2015quality,rosenthal2016quality}. In some cases, such as~\cite{clark1996sensory}, their introduction is motivated by mathematical convenience. In others, such as~\cite{rosenthal2015quality}, their introduction is closely tied to laboratory features, such as the topology of physical stimuli. Therefore, it is not so clear whether the prominent role metric structures play is due to an intimate relation to consciousness or just due to them being a very handy mathematical tool. A principled investigation, we take it, is highly imperative.

A metric structure $(M,d)$ consists of a set $M$ of `points' together with a function $d:M \times M \rightarrow \R$. Therefore, the domains of the structure are $M$ and $\R$. For the function to be a metric function, three axioms have to be satisfied for all points $x,y,z \in M$:
\begin{enumerate}
    \itemD \emph{Positive definiteness}, which requires that $d(x,y) \geq 0$ and that $d(x,y) = 0$ if and only if $x=y$.
    \itemD \emph{Symmetry}, which requires that $d(x,y) = d(y,x)$.
    \itemD The \emph{triangle inequality}, which requires that $d(x,y) \leq d(x,z) + d(z,y)$.
\end{enumerate}

We will turn to previous work on metric structure momentarily, but before doing so, we can ask whether this \emph{type} of structure could possibly be a structure of conscious experience, according to either \MDC\ or \MSC?

To answer this question, we look at one of the two domains of the metric structure, namely the real numbers $\R$. What is not well-known outside of mathematics is that the real numbers are not simply given, as the natural numbers or rationals might be, but that they have to be constructed in a comparably involved procedure. There are a small number of such procedures which yield equivalent results; the most common procedure is to construct the real numbers as equivalence classes of Cauchy sequences of rational numbers.%
    \footnote{A Cauchy sequence is, roughly speaking, a sequence of rational numbers that get arbitrarily close to each other. Two Cauchy sequences are defined to be equivalent if the difference between their elements tends to zero.}

If real numbers are equivalence classes of sequences of rational numbers, it is hard to see how one might reasonably claim that these correspond to aspects of conscious experiences, as required by both \MDC\ and \MSC. Independently of whether aspects are taken to denote qualities, qualia or phenomenal properties, there do not seem to be aspects that are equivalence classes of sequences of rational numbers. Phenomenal similarity, relative similarity, or similar aspects which have been associated to metric structures in previous work (see for example~\cite{clark1996sensory,lee2021modeling,rosenthal2010think}), do not have anything to do, in our eyes, with Cauchy sequences of rational numbers or any of the other real number construction schemes, such as Dedekind cuts.

Therefore, when understood in the precise sense of the term, a metric structure can neither be a structure of conscious experience~\MSC, nor a structure to describe conscious experience in the sense of~\MDC; it can only be an auxiliary tool.

The way to move forward, if one wants to consider something like a metric structure as a structure of conscious experience, is to restrict the target domain of the metric function $d$ to something that could reasonably be ``in'' conscious experience; that is, something closer to degrees of phenomenal similarity or a similar concept, for example, natural or rational numbers. We will now turn to this option. But it is important to note that most theorems about metric spaces cease to hold true if the real numbers are replaced by something else. That's because the convergence properties of sequences of real numbers (called completeness~\cite{rudin1970real}) are crucial for the definition of metric spaces.

The received wisdom on how to link metric structures to conscious experience is summarized concisely by Lee when discussing the ``standard framework'' for modeling mental qualities~\cite{lee2021modeling}:
    \begin{quote}
    ``There are three main desiderata when constructing a model in the standard
    framework. First, there should be one-to-one correspondence between points in the
    model and qualities in the targeted domain. Second, points that are more distant in
    the model should represent qualities that are less phenomenally similar to each
    other. Third, points should have distance zero just in case the qualities represented
    by those points are phenomenally identical.''~\cite[p.\,14]{lee2021modeling}
    \end{quote}

The first desideratum states that the metric structure contains a set of points $M$ that corresponds to the targeted domain; for example, qualities of a certain type. The third desideratum alludes to one of the three axioms of the metric structure, namely positive definiteness; it is implicitly understood that the other axioms should also hold. But what does the second desideratum mean? And how does it relate to a metric function that takes values in some number range?

To interpret the second desideratum, we have to understand how phenomenal similarity may relate to numbers; this will lead us to ``degrees'' of phenomenal similarity. To have a foundation on which to build this understanding, we will assume that phenomenal similarity can be described by a strict partial order $(M,<)$ as introduced in Section~\ref{sec:ex:relativeSimilarity}. This assumption is in line with Clark's proposal in~\cite{clark1996sensory} and~\cite{clark2000theory}, where a metric structure is based on relative similarity.

Given a strict partial order $(M,<)$ that describes phenomenal similarity (meaning that $x<y$ if and only if $x$ is less similar to some reference than $y$), a metric-like function $d$ on $M$ can be given by defining
$$
d(x,y) = \textrm{length}_<(x,y)\:,
$$
where $\textrm{length}_<(x,y)$ denotes the number of elements of the shortest path that connects $x$ and $y$.%
    \footnote{Formally, it is defined as
    $$
    \textrm{length}_<(x,y) = \begin{cases} 
    0 & \textrm{if } x = y \\
    \min |\{ x \rightarrow y \}| & \textrm{if } x < y\\
    \min |\{ y \rightarrow x \}| & \textrm{if } y < x\\
    \infty & \textrm{otherwise} \:,
    \end{cases}
    $$
    where $x \rightarrow y$ denotes a path along the strict partial order from $x$ to $y$, 
    $|\{ x \rightarrow y \}|$ denotes the number of elements in the path (with $x$ counted but $y$ not counted), and where the minimum selects the shortest path.}
This definition of $d(x,y)$ means that the metric function counts degrees of phenomenal similarity in-between $x$ and $y$, so that ``points that are more distant in the model [...] represent qualities that are less phenomenally similar to each other'' (ibid.), as required by the second desideratum.

As the reader can check, this definition of $d(x,y)$ indeed satisfies the three axioms of a metric function. Furthermore, it takes values in the natural numbers $\N$, so that the real-number problem is avoided. So, with $d: M \times M \rightarrow \N$ defined as above, is $(M,d)$ a structure of conscious experience?

By assumption, $M$ consists of aspects of conscious experiences: qualities of the targeted domain.
Furthermore, there is no obvious reason why a bounded subset of the natural numbers 
should not be aspects of conscious experiences of a suitable set $E$. Therefore, Condition~\ref{MSC:domains} might well be satisfied. 
So, whether or not $(M,d)$ is a structure of conscious experience \MSC, as compared to just a structure to describe conscious experiences \MDC, boils down to whether there is a $d$-aspect for $d$ as defined above. Let us check if there is one.

An aspect $a$ is a $d$-aspect if and only if every variation $v$ which does not preserve $d$ with respect to relata $b_1, ..., b_m$ changes $a$ relative to $b_1, ..., b_m$, and vice versa. The definition of relata in case of the function $d$ means that $m=3$ and $b_3 = d(b_1, b_2)$. Therefore, a variation $v$ doesn't preserve $d$ with respect to $b_1, b_2, b_3$ if either the $v(b_i)$ do not all exist or if $v(b_3) \neq d(v(b_1), v(b_2))$, which means that the variation changes the experience of the number $b_3 = d(b_1, b_2)$
in such a way that it does not agree any more with the experienced distance of $v(b_1)$ and $v(b_2)$. A variation $v$ preserves $d$ with respect to relata $b_1, b_2, b_3$, on the other hand, if $v(b_3) = d(v(b_1), v(b_2))$.
Therefore, what we're looking for is an aspect $a$ of some experience $e$ in which the identity
$b_3 = d(b_1,b_2)$ holds, that changes relative to $b_1, b_2, b_3$ if the above identity breaks, and that does not change relative to $b_1, b_2, b_3$ if the above identity holds true. This must be true of all relata $b_1, b_2, b_3$ of $d$, that is, of all experiences which exhibit aspects $b_1$, $b_2$ and $b_3$ such that $b_3 = d(b_1,b_2)$.

Taking into account that $b_3$ is the experience of a number, the only aspect which can satisfy these two conditions is the experience of $b_1$ and $b_2$ having distance $b_3$, or put differently: the aspect in question would have to be an experience of `having distance', which is instantiated relative to $b_1$, $b_2$ and the number $b_3$.
We do not think there is such an aspect. Aspects might be experienced as more or less similar, but we doubt that they are experienced as being a specific number apart, be that number natural or rational. Therefore, even if watered down to avoid the issues with real numbers, a metric structure does not seem to be a structure of conscious experience.

So, in summary, there is one immediate and one deep reason why conscious experience does not have a metric structure. The immediate reason is that the real numbers with their very involved construction scheme do not seem to have anything to do with aspects of conscious experience. The deep reason is that we simply do not experience qualities or other aspects as being a particular number of degrees of similarity apart. This might explain why in approaches such as that of Clark~\cite{clark1996sensory,clark2000theory}, the details of the introduction of a metric structure point outside the realm of conscious experience.

\section{Phenomenal Unity and Topological Structure}\label{sec:ex:topological}

Our final example concerns topological structure. Interestingly, this is intimately tied
to phenomenal unity, the thesis that phenomenal states of a subject at a given time are unified~\cite{chalmers2003unity}.

Recall that we have introduced the set $A(e)$ to denote aspects of the conscious experience~$e$, where we have used the term `aspect' as a placeholder for concepts like qualia, qualities, or (instantiated) phenomenal properties.
Most examples of these concepts are ``independent'' from the experience in which they occur; they could be experienced together with a largely different set of aspects in a  different experience. Yet, experiences seem unified; their aspects are experienced as tied together in some essential way. This raises the question of what underlies this experience of the \emph{unity of a conscious experience}? As we will see, somewhat surprisingly, the answer is: a topological structure of conscious experience.

Much has been written about the question of phenomenal unity in the literature, for example~\cite{bayne2012unity,chalmers2003unity,cleeremans2003unity,mason2021model,roelofs2016unity,wiese2019what}, and in order to make use of some of the results, we assume that the term `aspect' denotes an instantiated phenomenal property or quale. The set of aspects $A(e)$, then, comprises the phenomenal properties or qualia which are instantiated in the experience~$e$, also called the \emph{phenomenal states} of the experience~$e$.%
    \footnote{A \emph{phenomenal state} is an instantiation of a phenomenal property, or quale, by a subject at a given time. This instantiation constitutes part of the experience of the subject at the time. An experience $e$, in our terminology, is an experience of a subject at a given time. Hence, a phenomenal state is an instantiation of a phenomenal property, or quale, in an experience $e$.}
Our question, then, is what it means that ``any set of phenomenal states of a subject at a time is phenomenally unified''~\cite[p.\,12]{chalmers2003unity}.

There are various answers one might give to this question. A promising answer is the so-called \emph{subsumptive unity thesis}, developed in~\cite{chalmers2003unity}:
    \begin{quote}
    ``For any set of phenomenal states of a subject at a time,
    the subject has a phenomenal state that subsumes each of the states in that set.''~\cite[p.\,20]{chalmers2003unity}
    \end{quote}
According to this thesis, what underlies the experience of the unity of a conscious experience is that for any set~$X$ of phenomenal states in the conscious experience, there is a further phenomenal state that subsumes each of the states in $X$. This phenomenal state characterizes what it is like to be in all of the states of $X$ at once~\cite[p.\,20]{chalmers2003unity}.

Put in terms of aspects, the subsumptive unity thesis says that for any set $X \subset A(e)$ of aspects of an experience, there is an additional aspect in $A(e)$ that subsumes the aspects in~$X$.
This aspect is the experience of what it is like to experience the aspects in $X$ as part of one experience $e$ together, the experience that they are \emph{unified}, as we will say. Let us call this aspect the \emph{phenomenal unity aspect} of $X$ and denote it by $a_X$. It is instantiated relative to the elements of $X$.

Phenomenal unity gives rise to a mathematical structure of conscious experience. To see how, let us use the symbol $\T$ to denote a collection of subsets of $A(e)$, to be specified in more detail below. Every subset of $A(e)$ is a unary relation on $A(e)$,%
    \footnote{An $m$-ary relation on a set $X$ is a subset $R$ of $X^m$. Hence, a unary relation, where $m=1$, is a subset of $X$.}
and hence also on the set~$\A$ that comprises all aspects of the experiences in $E$.
Therefore, $(\A,\T)$ is a mathematical structure; it has domain $\A$ and its structures are the unary relations in~$\T$.
The next paragraph shows that because of the subsumptive unity thesis, the mathematical structure $(\A,\T)$ is a mathematical structure of conscious experience according to~\MSC.

Because $\A$ is the set of all aspects of $E$, Condition~\ref{MSC:domains} of \MSC\ is satisfied. Therefore, only Condition~\ref{MSC:aspect}\ remains to be checked. This condition is satisfied because for every set $X \in \T$, the phenomenal unity aspect $a_X$ is an $X$-aspect. To show that this is the case, we need to check that a variation does not preserve $X$ with respect to relata $b_1, ... , b_m$ if and only if it changes $a_X$ relative to $b_1, ..., b_m$.
Let $v:A(e) \rightarrow A(e')$ be a variation that does not preserve $X$ with respect to relata $b_1, ..., b_m \,$. The relata of the subset $X$ are the elements of that subset. Therefore, we have $b_1, ..., b_m \in A(e)$, so that the subsumptive unity thesis implies that there is a phenomenal unity aspect $a_X$ relative to the $b_1, ..., b_m$ in $A(e)$. The condition that $v$ does not preserve $X$ furthermore implies that either not all of the $v(b_i)$ exist or that at least one of them is not in the set $X$. Therefore, there is no phenomenal unity aspect $a_X$ relative to $v(b_1), ..., v(b_m)$ in $A(e')$. Hence, the variation $v$ changes $a_X$ relative to $b_1, ..., b_m \in X$.
Vice versa, let $v:A(e) \rightarrow A(e')$ be a variation which preserves $X$ with respect to relata $b_1, ..., b_m$. This implies that $a_X$ is instantiated relative to $b_1, ..., b_m$ in $A(e)$. The condition that $v$ preserves $X$ furthermore implies that $v(b_1), ..., v(b_m)$ exist, and that they are elements of~$X$. Therefore, $a_X$ is also instantiated relative to $v(b_1), ..., v(b_m)$ in $A(e')$. This shows that the variation does not change $a_X$ relative to $b_1, ..., b_m$.
Thus, $a_X$ is indeed an $X$-aspect. And because that is true for any $X \in \T$, $(\A,\T)$ indeed satisfies Condition~\ref{MSC:aspect} and hence~\MSC.

The previous paragraph proves that, if the subsumptive unity thesis holds true for all sets $X$ in $\T$, then $(\A,\T)$ is indeed a mathematical structure of conscious experience. As we will explain next, this structure is intimately tied to a topological structure.

A topological structure $(M,\T)$ consists of a set $M$ and a collection $\T$ of subsets of $M$. The collection has to satisfy three axioms, and there are a few different ways of formulating these axioms. Here, we choose the formulation that corresponds to what is usually called `closed sets'. The axioms are:
\begin{enumerate}
    \itemD The empty set $\emptyset$ and the whole set $M$ are both in $\T$.
    \itemD The intersection of any collection of sets of $\T$ is also in $\T$.
    \itemD The union of any finite number of sets of $\T$ is also in $\T$.
\end{enumerate}
Having specified what a topological structure is, we return to the structure $(\A,\T)$, which is induced by phenomenal unity, and ask what the collection $\T$ of subsets is?

First, it is important to note that the subsumptive unity thesis does not provide a phenomenal unity aspect $a_X$ for every subset of $\A$; it can only provide such an aspect for a set of aspects that are actually experienced together, that is, for a subset $X$ of $A(e)$. Therefore, $\T$ is not the discrete topology introduced in Section~\ref{sec:previous}. 
Second, it also cannot be the case that it provides a phenomenal unity aspect for every subset of $A(e)$. That's because then there would be an infinite regress: for every subset $X$ of $A(e)$ there would be a new aspect $a_X$ in $A(e)$, giving a new subset $X \cup \{a_X\}$ that would give a new phenomenal unity aspect $a_{X \cup \{a_X\}}$, and so forth. This problem is well-known in the literature~\cite{bayne2005divided,wiese2019what}.
Rather, we take it, the quantifier `any set' in the subsumptive unity thesis must be understood as `any set of aspects that are experienced as being unified'. While it is arguably the case that the whole set of aspects $A(e)$ of an experience is always experienced as unified, introspection suggests that we consciously experience only a select group of aspects as unified at a time.%
    \footnote{This solves the infinite regress problem because, arguably, we do not always experience the phenomenal unity aspects as unified with the sets they correspond to. So, there is not always a phenomenal unity aspect $a_{X \cup \{a_X\}}$ for the set that consists of $a_X$ and $X$.}

So, which sets of aspects do we experience as unified? While it might be difficult to give a general answer to this question, there is a special case where a sufficiently detailed specification can be given: the case of regions in visual experience. Here, `regions' are sets of positions of the space that visually perceived objects occupy.%
    \footnote{
    It is also plausible to think that visual experiences do not contain positions as aspects, but only regions. However, assessing whether or not this is the case goes beyond the scope of this paper. Here, we assume that positions are aspects of visual experiences.
    }
The positions in a region are experienced as unified. Therefore, the regions of visual experience are members of the collection $\T$ which is induced by phenomenal unity. 
Furthermore, they appear to satisfy the axioms of a topology as stated above: the whole set of positions in a visual experience is a region; it seems to be the case that intersections of regions in visual experience are also regions in visual experience; and
it seems to be the case that the union of any two regions in visual experience is also a region in visual experience.
For the empty set, no $S$-aspect of consciousness is required (there are no relata of the corresponding unary relation), so we can take the empty set to be a member of~$\T$. Thus, all axioms of a topology are satisfied.

Therefore, if we take $M$ to denote the position aspects of visual experiences, and choose $\T$ to comprise the regions of visual experience, then $(M,\T)$ is indeed a topological structure. And, as shown above, it is a structure of conscious experience as defined in~\MSC. 
We thus find that, because of the subsumptive unity thesis, this topological structure is indeed a mathematical structure of conscious experience; much like conjectured in~\cite{tallon2022topological}, it is a topology of the visual content of subjective experience.

\section{The Three Problems Revisited}\label{sec:threeProbsAgain}

In this section, we discuss how the new approach~\MSC, which we have developed in Section~\ref{sec:mathematicalStructure}, resolves the three problems discovered in Section~\ref{sec:previous}.

\subsection*{Problem 1: Incompatible Structures}
The first problem was that the condition~\MDC, which has been applied in previous approaches, admits incompatible structures to conscious experience. Is this also true of~\MSC?

If two structures are incompatible, then there exists at least one automorphism of one structure that is not an automorphism of the other structure.%
    \footnote{
    Automorphisms are structure-preserving mappings from a structure to itself. 
    Put in terms of the terminology we have introduced in Section~\ref{sec:mathematicalStructure}, automorphisms are mappings $v$ that map the domains of a structure to themselves. These mappings have to be bijective, and they have to preserve the structure, meaning that they have to satisfy~\ref{preserves:relation} for all elements of the domain in case of relations, and~\ref{preserves:function} for elements of the domains in the case of functions.
    }
As we explain below, this condition implies that two incompatible structures cannot have an $S$-aspect in common. Therefore, it is not possible for two incompatible structures to pertain to conscious experience in the exact same way; so, \MSC\ indeed resolves the problem of incompatible structures. 

Let $S$ and $S'$ denote two incompatible structures with the same domains.
Then, there is at least one automorphism of one structure that is not an automorphism of the other structure. Let us denote such an automorphism by $v$ and assume that it is an automorphism of $S$ but not of $S'$.
Because $v$ is not an automorphism of~$S'$, it follows that there is at least one set of relata $b_1, ..., b_m$ of $S'$ in some $A(e)$, such that the variation $v:A(e) \rightarrow A(e)$ induced by the automorphism does not preserve $S'$ with respect to these relata. On the other hand, because $v$ is an automorphism of $S$, it follows that this variation preserves $S$ with respect to $b_1, ..., b_m$.
If an aspect $a$ is an $S'$-aspect, then, applying the definition of $S'$-aspects, we find that the variation $v$ needs to change it. In contrast, if an aspect $a$ is an $S$-aspect, then, applying the definition of $S$-aspects, we find that the variation $v$ needs to not change it; either because the $b_1, ..., b_m$ do not constitute relata of $S$, or because the variation $v$ preserves $S$ with respect to relata $b_1, ..., b_m$. So, because an aspect cannot be both changed and not changed under a single variation, there cannot be an aspect $a$ that is both an $S$-aspect and an $S'$-aspect.

\subsection*{Problem 2: Arbitrary Re-Definitions.}
The definition~\MSC\ also resolves the problem of arbitrary re-definitions. That's because any re-definition changes the relata of the respective structure, and therefore generates an own, independent condition for something to be an $S$-aspect of the redefined structure. Whether or not this new $S$-aspect is a part of conscious experience is a substantive question that depends on the actual experiences of the subject under consideration; it is not automatically the case.

Consider, as examples, the cases of rescaling a metric, which we have introduced in Section~\ref{sec:previous}. If, per assumption, $(M,d)$ were a structure of conscious experience, then for any relata $(b_1,b_2,d(b_1,b_2))$, the condition for $d$-aspects would have to be satisfied. Rescaling this to $(M,C \cdot d)$ generates a new condition because now, the relata to be considered are $(b_1,b_2,C \cdot d(b_1,b_2))$. These are different relata, and correspondingly, different experiences and different variations will enter the definition of a $C \cdot d$-aspect.
The same is true for an $(f(a)+f(b)) \cdot d(a,b)$-aspect. Whether or not these structures satisfy~\MSC\ depends on the details of the conscious experiences under consideration; but they do not automatically satisfy~\MSC\ just because $(M,d)$ does.

\subsection*{Problem 3: Indifference to Consciousness.}
The third problem is resolved, finally, because of the introduction of $S$-aspects, which are a counterpart ``in'' conscious experience to the proposed mathematical structure.
Because $S$-aspects are part of the definition~\MSC, any application of \MSC\ requires one to engage with details of the conscious experiences of the subject under consideration; \MSC\ is not indifferent to conscious experience.

Consider, for example, the two topological structures of Section~\ref{sec:previous}. While \MDC\ only required us to check whether the structures are about aspects and satisfy the axioms, \MSC\ also requires us to check whether there is an $S$-aspect in conscious experience that corresponds to the structures.
As we have seen in Section~\ref{sec:ex:topological}, this involves a careful investigation of conscious experience, for example, concerning the phenomenal unity.

\section{Conclusion}\label{sec:conclusion}

In this article, we investigated mathematical structures of conscious experience.
Our main result is a definition of what \emph{mathematical structures of conscious experience} are. This definition provides a general method to identify and study structures of conscious experience; it is grounded in a foundational understanding of mathematical structures as laid out by mathematical logic; and it provides a link between the abstract formal entities of mathematics, on the one hand, and the concreta of conscious experience, on the other hand, see Section~\ref{sec:proposal}. Our definition also resolves three problems that interfere with recent approaches that relate mathematical structures to conscious experience, see Section~\ref{sec:previous}.

What we consider noteworthy about our definition is that it is \emph{conceptually neutral}, meaning that it does not rely on any specific conception of conscious experience or aspects. Rather, it is applicable to any conception of `conscious experience' and `aspects' in which every conscious experience comes with a set of aspects. This includes common conceptions built on qualities, qualia, or phenomenal properties, but also less common ideas built on atomistic conceptions of states of consciousness or phenomenal distinctions.
Furthermore, our definition is \emph{methodologically neutral}, meaning that it can be combined with many methods, practices, and procedures that are used to investigate conscious experience, spanning empirical, analytical, and phenomenological research.
That is because the definition rests on the concept of variations, and variations can be induced in three major ways: introspectively (for example, as in Husserl's imaginary variations~\cite{husserl1936crisis}); in a laboratory by change of stimuli; or theoretically based on a proposed theory of consciousness.

As first applications of our definition, we considered \emph{relative similarity}, \emph{metric spaces}, and \emph{topological spaces}. We found that relative similarity, which plays an important role in several constructions of quality spaces, is indeed a mathematical structure of conscious experience, see Section~\ref{sec:ex:relativeSimilarity}. Topological spaces are too, but for a surprising reason: they are intimately related to phenomenal unity, see Section~\ref{sec:ex:topological}. 
Metric spaces, however, are not structures of conscious experience; see Section~\ref{sec:ex:metric}. One of the two reasons for that is that metric spaces are built on the real numbers, which have to be constructed with involved procedures using, for example, equivalence classes of Cauchy sequences that do not seem to correspond to aspects of conscious experience, however conceived. 

We view the result presented here as one further step in a long journey to investigate conscious experience mathematically. This step raises new questions and creates new opportunities, 
both of which can only be explored in an interdisciplinary manner. A new question, for example, is 
 whether our result on mathematical structures might open new perspectives on measurements of consciousness~\cite{irvine2013measures}, as arguably promised by the Representational Theory of Measurement~\cite{krantz1971foundations} whenever an axiomatic structure on a target domain is available. 
A new opportunity, in our eyes, is the unification of the various approaches to represent consciousness mathematically that have emerged in different fields. Because our result is conceptually and methodologically neutral, it can provide a framework for this unification; but to develop a unification that is both useful and valuable in practice requires experts from the respective fields. 
We hope that, ultimately, our result provides a basis for developing a common formal language to study consciousness.

\subsection*{Acknowledgments} 
We would like to thank the participants of the \emph{2022 Modelling Consciousness Workshop} and of the \emph{Models of Consciousness 3} conference, both organized under the umbrella of the Association for Mathematical Consciousness Science, as well as members of the \emph{Munich Center for Mathematical Philosophy} for fruitful discussions and helpful comments, and in particular Jonathan Mason for valuable feedback on the manuscript.
This research was supported by grant number FQXi-RFP-CPW-2018 from the Foundational Questions Institute and Fetzer Franklin Fund, a donor advised fund of the Silicon Valley Community Foundation. We would like to thank the Dutch Research Council (NWO) for (partly) financing TL's work on project number 182.069 of the research programme Fluid Spintronics, and the Mathematical Institute of the University of Oxford for hosting JK while working on this project.


\end{document}